\documentclass[aps,prb,twocolumn,showpacs,preprintnumbers,amsmath,amssymb,superscriptaddress]{revtex4}

\usepackage{graphicx}
\usepackage{dcolumn}
\usepackage{bm}
\newcommand{\nc}{\newcommand}
\nc{\be}{\begin{equation}}
\nc{\ee}{\end{equation}}
\nc{\bea}{\begin{eqnarray}}
\nc{\eea}{\end{eqnarray}}
\nc{\bean}{\begin{eqnarray*}}
\nc{\eean}{\end{eqnarray*}}
\nc{\mb}{\mbox}
\nc{\rnc}{\renewcommand}
\nc{\vk}{\mb{\bf k}}
\nc{\vp}{\mb{\bf p}}
\nc{\vn}{\mb{\bf n}}
\nc{\vq}{\mb{\bf q}}
\nc{\rr}{\mb{\bf r}}
\nc{\vz}{\hat {\mb{\bf z}}}
\nc{\vj}{\mb{\boldmath$j$}}
\nc{\vg}{\mb{\boldmath$g$}}
\nc{\x}{\mb{\boldmath$x$}}
\nc{\A}{\mb{\boldmath$A$}}
\nc{\va}{\mb{\boldmath$a$}}
\nc{\vs}{\mb{\boldmath$\sigma$}}
\nc{\vpi}{\mb{\boldmath$\pi$}}
\nc{\nab}{\nabla}
\nc{\X}{\sf x}

\begin{document}


\title{Dependence of the intrinsic spin Hall effect on spin-orbit
interaction character}
\author{K. Nomura}
\affiliation{Department of Physics, University of Texas at Austin,
Austin TX 78712-1081, USA}
\author{Jairo Sinova}
\affiliation{Department of Physics, Texas A\&M University, College
Station, TX 77843-4242, USA}
\author{N. A. Sinitsyn}
\affiliation{Department of Physics, University of Texas at Austin,
Austin TX 78712-1081, USA}
\author{A. H. MacDonald}
\affiliation{Department of Physics, University of Texas at Austin,
Austin TX 78712-1081, USA}

\date{\today}

\begin{abstract}

We report on a comparative numerical study of the spin Hall conductivity
in two-dimensions for three different spin-orbit interaction models; the 
standard k-linear Rashba model, the k-cubic Rashba model that describes
two-dimensional hole systems, and a modified k-linear Rashba model in 
which the spin-orbit coupling strength is energy dependent. 
Numerical finite-size Kubo formula results indicate that the spin Hall
conductivity of the k-linear Rashba model vanishes for frequency $\omega$ 
much smaller than the scattering rate $\tau^{-1}$, with order one relative fluctuations
surviving out to large system sizes. 
For the k-cubic Rashba model case, the spin Hall conductivity does not 
depend noticeably on $\omega \tau$ and is finite in the {\em dc} limit, in agreement with experiment.
For the modified k-linear Rashba model the spin Hall conductivity 
is noticeably $\omega \tau$ dependent but approaches a finite value in the {\em dc} limit.
We discuss these results in the light of a spectral decomposition of the spin Hall 
conductivity and associated sum rules, and in relation to a proposed separation of the spin Hall conductivity 
into skew-scattering, intrinsic, and interband vertex correction contributions. 

\end{abstract}

\maketitle

\noindent

\section{Introduction}
Interest in spintronics\cite{wolf,parkin,awschalom,dassarma,schiffer,dietl} has been heightened 
by the technological impact of ferromagnetic metal based devices and by ferromagnetic 
semiconductor materials advances.  Theoretical attention has recently focused on
spintronics effects in paramagnetic materials, and in particular on the spin Hall
effect\cite{dyakonov} in which an electric field induces a transverse spin current.
Murakami {\it et al.} \cite{murakami} and Sinova {\it et al.} \cite{sinova} have
argued in different contexts that the spin Hall conductivity can be dominated
by a contribution that follows from the distortion of Bloch electrons
by an electric field and therefore approaches an intrinsic value in the
clean limit.  The intrinsic spin Hall conductivity adds to the skew scattering 
contribution that had been the focus of earlier theoretical work,\cite{dyakonov,hirsch,zhang}
and can be altered by disorder vertex corrections.

\begin{figure}[h]
\begin{center}
\includegraphics[width=0.45\textwidth]{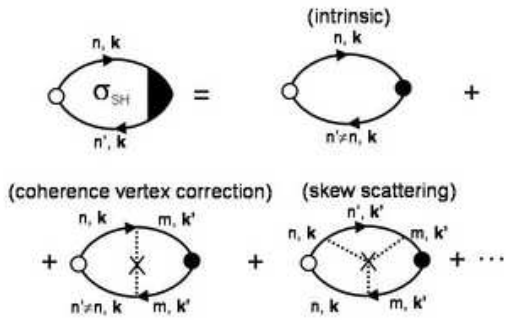}
\caption{Schematic diagrammatic representation of intrinsic, interband coherence vertex correction, and 
skew scattering contributions to the spin Hall conductivity.  The total spin current (left vertex - open circle corresponding to ${\bf j}^z=\{{\bf v},s_z\}/2$) 
induced by an electric field (right vertex - dark circle corresponding to ${\bf j}= -e{\bf v}$) can be separated into a contribution from the density-matrix response 
that is off-diagonal in band index and a contribution from the density-matrix response that is diagonal 
in band index.  The diagonal response is dominated by a skew scattering contribution that is in general proportional
to the Bloch electron scattering time $\tau$.  For the models studied here, the spin-current operator is purely
off diagonal in band index and the skew scattering contribution is absent.
The off-diagonal response has a leading contribution, the intrinsic spin Hall
conductivity, that is completely independent of disorder and is property of the host Bloch bands.
Vertex corrections to the intrinsic spin-Hall conductivity are also independent of $\tau$ 
in the limit of weak scattering, but depend on both band-structure and 
disorder potential.  The solid lines in the figure represent Born approximation
Greens functions.}
\label{diagrams}
\end{center}
\end{figure}
 
The proposed intrinsic spin Hall conductivity has drawn theoretical
attention \cite{culcer,inoue,halperin,khaetskii,schwab,dimitrova,rashba1,rashba2,loss}
to this unfamiliar transport coefficient. It has been argued 
that the intrinsic spin Hall conductivity does not survive in the diffusive transport thermodynamic limit, 
either generally or for the specific case of the two-dimensional electron system with 
Rashba spin-orbit interactions (R2DES)\cite{rashba0} studied by Sinova {\it et al.} \cite{sinova}. 
Several researchers have pointed out that in the Rashba 2DEG case, ladder vertex 
disorder corrections~\cite{inoue,schwab,loss} in the Kubo formula 
lead to a vanishing spin Hall current.  The same conclusion has 
been reached via quantum Boltzmann theory\cite{halperin,khaetskii} calculations which 
capture the same physics. 
These vertex correction claim is specific to the R2DES case, and has
specifically been discounted for two-dimensional hole gases~\cite{schliemann,bernevig}, p-doped
bulk semiconductors and the modified Rashba coupling case.~\cite{murakami2} 
Several numerical studies basing on the Landuer-Buttiker approach in
mesoscopic systems with leads also have been performed, but lack of clear 
trends in the size dependence of the spin Hall conductances they evaluate\cite{sheng,nikolic,hankiewicz}, 
and the possibility of edge effects near the contacts, cannot be
connected to
the possibility of the spin Hall effect in the thermodynamic limit.

In this article we approach these issues numerically by evaluating
the finite-size Kubo formula\cite{thouless,nomura,sheng2} for the spin Hall conductivity for 
two-dimensional electrons with three different spin-orbit coupling models.
The motivation for the models we have chosen follows from the parsing of 
spin-Hall conductivity contributions illustrated schematically in Fig.~\ref{diagrams}. 
Quite generally the charge and spin-current operators are diagonal in
Bloch wavevector in any model with a spin-independent impurity potential,
but have matrix elements that are both diagonal and off-diagonal in band index.  (The models we study
have only two-bands and it is usually convenient to consider them as representing the two
spin states of a spin-$1/2$ particle.)  The presence of diagonal matrix elements means 
that current will not decay in a perfect crystal and the longitudinal conductivity 
(and Hall conductivity) is consequently limited by disorder scattering of Bloch states.
 
We start from a linear response Kubo formula approach:\cite{mahan}
\bea
 \sigma_{\mu\nu}^{\alpha}(\omega)&=&-\frac{K_{\mu\nu}^{\alpha}(\omega)-K_{\mu\nu}^{\alpha}(0)}{i\omega}
\eea
where 
\bea
K_{\mu\nu}^{\alpha}(\omega+i\eta)&=&-\frac{i}{\hbar\Omega}\int_0^{\infty}dt\langle[j_{\mu}^{\alpha}(t),j_{\nu}]\rangle e^{i(\omega+i\eta)t},
\label{eq:kernel} 
\eea
$j_{\mu}^{\alpha}$ is the $\alpha$ component spin or charge current
in $\mu$ direction, and $\Omega$ is the volume of the system.
When disorder is treated perturbatively, the standard ladder diagram 
approximation expresses the kernel $K$
in Eq.~(\ref{eq:kernel}) in the following form.\cite{mahan}:
\bea
 K_{\mu\nu}^{\alpha}(i\nu_m)=\frac{1}{\beta\Omega}\sum_{\vk,\zeta_n} {\rm
 tr}[j_{\mu}^\alpha
G(\vk,\zeta_n+\nu_m)J_{\nu}G(\vk,\zeta_n)],
\eea
where $G$ is a Matsubara formalism Born-approximation Greens function,
$J_{\nu}$ is the current operator renormalized by ladder-diagram vertex corrections,
and all four quantities are matrices in certain bases. 
In this approximation the longitudinal conductivity ends up being dominated by
terms that are diagonal in band index at each current vertex and end up being 
proportional to the Bloch state scattering time $\tau$.   This contribution captures the Boltzmann theory 
physics in which the current is due to field induced changes in the occupation probabilities of Bloch states.
It turns out that the spin Hall conductivity does in general have a corresponding contribution, 
but only if scattering violates the principle of 
microscopic reversibility, {\em i.e.} only if the $a \to b$ scattering rate is 
not equal to $b \to a$ scattering rate where $a$ and $b$ are Bloch states.  Indeed this property
is violated when scattering amplitudes are evaluated beyond the Born approximation, 
as illustrated schematically in Fig.~\ref{diagrams}, 
and the spin Hall conductivity ends up being proportional to $\tau \times S$, where the 
{\em skewness} $S$ is a measure of the violation of microscopic reversibility.
\cite{skew}
The terminology we use here is borrowed from the theory of the anomalous Hall effect in ferromagnetic 
metals and semiconductors,
which is strongly analogous to the spin Hall effect in paramagnetic metals and semiconductors.
We identify the contribution to the spin Hall conductivity that is proportional to $\tau$ and 
due to changes in Bloch state occupation probabilities, {\em i.e.} to a response that is diagonal in 
both wavevector and band indices, as the {\em skew scattering}
contribution.\cite{dyakonov,hirsch,zhang} 
For the models we study here the spin-current operator is purely off diagonal in band index,
as we explain explicitly below.  It follows that {\em the skew scattering contribution to the 
spin-Hall conductivity vanishes for the models we study}.

We define the intrinsic spin Hall conductivity as the interband spin Hall conductivity $\sigma(\omega)$ 
of a disorder free system, which has a finite value in general in the $\omega \to 0$ {\em dc} limit.
This quantity is a property of the band structure of the perfect crystal, hence the term
intrinsic.  The {\em dc} limit of the interband conductivity can, however, be altered by 
disorder even in the limit of arbitrarily weak disorder, $\tau \to \infty$ as illustrated 
schematically in Fig.~\ref{diagrams}.  In perturbation theory the factors of $\tau^{-1}$ associated 
with Born approximation disorder scattering can be canceled by factors of $\tau$ associated with 
products of two-Greens functions that have the same band index. The evaluation of the 
disorder correction to the interband spin-Hall conductivity requires the evaluation
of a ladder sum.  The renormalized current with the ladder correction is given by solving
the vertex equation,
\bea
&&{\bf J}(\vk;z,z')=\vj(\vk)\nonumber \\ &&
 \ \ +\sum_{\vk'}\frac{|V(\vk-\vk')|^2}{\Omega^2}
G(\vk',z){\bf J}(\vk';z,z')G(\vk',z').
\eea
  Explicit evaluation of this ladder sum correction requires some approximations and can
normally be accomplished only for very simple disorder models, or in the limit of small 
spin-orbit interactions.  These limitations of perturbation theory motivate the numerical study
reported on here. 

We have reported previously on the influence of disorder on the spin Hall conductivity 
of the k-linear Rashba model, concluding that it remains finite in the thermodynamic 
limit.\cite{nomura}  This conclusion is at odds with our current numerical findings, extrapolating 
to infinite system size and then to zero frequency, which are 
consistent with the perturbation theory conclusion that the dc spin Hall conductivity 
of this model is zero.    
The numerical studies are complicated by the strong frequency
dependence and large fluctuations in the spin Hall conductivity
that occur in finite-size calculations.  The earlier calculations erred by using the 
frequency dependence of the longitudinal conductivity, which has corrections that 
vary like $(\omega \tau)^2$ compared to the $(\omega \tau)^{1}$ dependence of the spin 
Hall conductivity discussed below, to judge whether or not the {\em dc} limit has been
reached.   
The new findings supersede the conclusions reached in 
Ref. [\onlinecite{nomura}] with regards to the thermodynamic {\em dc} limit. 
Our numerical results for the k-linear and k-cubic Rashba models are 
now consistent with analytic calculations 
which consider only the thermodynamic limit within the diffusive regime, 
hence resolving the controversy that has been associated with 
the linear Rashba model.\cite{note_on_Murakami}

In the k-cubic Rashba model, which approximately describes two-dimensional 
valence band holes in a narrow quantum well with structural inversion asymmetry\cite{schliemann,winkler},
the spin Hall conductivity does not show measurable $\omega \tau$ dependence.
For this model, its $\omega \to 0$ limit is consistent with the pure intrinsic value.
This observation suggests that the spin Hall induced edge spin accumulations\cite{awschalomscience} 
recently seen in two-dimensional hole systems\cite{wunderlich} follows from the 
intrinsic spin Hall effect.
We also study a modified Rashba interaction which combines elements of the 
k-linear and k-cubic Rashba model and provides an approximate 
model for conduction band quantum well states in inverted gap materials 
like HgTe.\cite{HgTe} As we explain below the modified model has spin-orbit 
splitting that varies like $k^3$, as in the k-cubic Rashba model, but a wavevector 
dependent Zeeman field whose in-plane orientation rotates once when
the wavevector rotates once around the Fermi surface as in the k-linear
Rashba model.  In perturbation theory,
the second property implies that an angular integral that appears in the 
vertex correction calculation and (for short-range impurity scattering)
vanishes in the k-cubic Rashba model, is non-zero.
Vertex corrections to the intrinsic spin Hall conductivity survive for the 
modified model.  Our numerical results demonstrate, however, that the corrections 
are present but do not cause the total spin Hall conductivity to vanish as it does
for the k-linear Rashba model.  The special property of the k-linear Rashba model
that causes the spin Hall conductivity to vanish is related to the equation of motion of 
the spin-operator.\cite{dimitrova,loss}

Our paper is organized as follows.  In Section II we describe the spin-orbit coupling and disorder terms in the 
model we study numerically.  The spin-orbit interaction can be described in terms of a position and momentum 
dependent Zeeman field, whose orientation variation as a function of wavevector plays the key role in 
spin Hall conductivity calculations for these models. The disorder model we employ assumes a scalar random potential.  
We argue that as long as the random potential is dominantly spin-independent, this assumption is not essential.
In Section III we introduce the finite-size-system Kubo formula which expresses the spin Hall conductivity in terms of Hamiltonian
eigenstates of a finite size two-dimensional electron system with area $L^2$ and periodic 
boundary conditions.  The spin Hall conductivity
evaluated using this formula tends to fluctuate wildly from disorder realization to disorder realization and 
is very sensitive to avoided level crossings that occur close to the Fermi energy.  These fluctuations are 
conveniently mitigated by evaluating 
the spin Hall conductivity for a continued complex frequency $\omega \to 
z = i \eta$ along the imaginary axis.  The {\em dc} spin Hall conductivity should be evaluated by 
first letting $L^2 \to \infty$ and then $\eta \to 0$.  Our expectation is that for systems much larger 
than a mean-free-path in size, $L^2$ dependence will appear only for $\eta$ smaller than or 
comparable to the finite size level spacing $\delta E$.   Thus we should be able to extrapolate to the 
{\em dc} value as long as systems sizes can be reached numerically 
that are large enough to make other characteristic 
energy scales like the spin-orbit splitting and the life-time broadening energy $\hbar/ \tau$, much larger
than $\eta \gg \delta E$.  In Sec. IV we present our numerical results for the finite-size Kubo formula and 
discuss its extrapolation to infinite system sizes.
In Section V we discuss a spectral representation for the spin Hall effect and some associated 
sum rules.  In Sec. VI we discuss the equation of motion of the spin operator for the k-linear Rashba model.
Using the fact that the time derivative of the spin operator is proportional to the spin Hall current for
this model, we are able to demonstrate that the finite size spin-Hall conductivity 
is always zero when averaged over boundary conditions. The typical size
of the spin-Hall conductivity fluctuation in a given finite size system, is however,
much larger than the intrinsic spin Hall conductivity.  Finally in Sec. VII we present our conclusions. 

\noindent
\section{Model Hamiltonian}
We base our studies on a series of models with generalized Rashba spin-orbit interaction
of the form:
\bea
H={\hbar^2 {\vk}^2}/{2m}+i\lambda g(|\vk|)(k_-^\alpha \sigma_+-k_+^\alpha \sigma_-)
\eea
where $m$ is the effective carrier mass, $k_{\pm}=k_x\pm i k_y$, and $g$
is either unity or a function of $|\vk|$. (We set
$\hbar=1$ for simplicity.)  The Rashba model for an inversion 
asymmetric conduction band quantum well
is generated by choosing $\alpha=1$ and $g=1$. 
Valence band quantum wells have a more complex
structure.\cite{wunderlich} In the thin quantum well limit, light hole bands become
energetically irrelevant and the heavy hole bands can be effectively described by Eq.(1) with
$\alpha=3$, the so called k-cubic Rashba model.
We study in addition a modified Rashba model with 
$g=k^2$ and $\alpha=1$ ($i.e.$ $H_R=\lambda k^2[\vk\times\vz]\cdot\vs$),
which provides an approximate model for conduction electron quantum well states in inverted gap
materials.\cite{HgTe}
In the following we use the
Fermi energy $E_F$ and the inverse Fermi wave number $k_F^{-1}$ in the absence of 
both spin-orbit coupling and disorder as the units of energy and length respectively.
We take a disorder model consisting of uncorrelated short-range scalar impurity potentials: 
$V(\rr)=\sum_{I=1}^{N_i} V\delta(\rr-{\bf R}_I)$
 which satisfies 
$\langle V(\vq) V(\vq')^* \rangle=N_iV^2\delta(\vq-\vq'),$
 where $V(\vq)$ is the Fourier component of $V(\rr)$ and $N_i=n_iL^2$ is the number of impurity 
scatterers which we take to correspond roughly to the number of carriers. We choose this type of disorder potential
model, rather than the more realistic finite correlation length model utilized in previous studies \cite{nomura}
in order to connect more directly with the analytical results. 

We diagonalize the finite-size disordered electron Hamiltonian in the $\lambda=0$ eigenstate
basis and introduce a hard cutoff at a sufficiently large momentum $\Lambda$. (This means, of course, that 
the disorder potential has an effective correlation length $\sim \Lambda^{-1}$.)  Our calculations are 
performed at a fixed carrier density $n_e=k_F^2/2\pi$ and at finite system sizes (see below) up to 
$70 k_F^{-1}$,
larger than the mean free path $\sim 10k_F^{-1}$ and the Fermi wavelength 
$\sim k_F^{-1}$. 

Finally we introduce the charge and spin current operators.
The charge current definition, $\vj=\pm e{\bf v}=-
\partial H/\partial {\bf A}$ follows from the charge conservation 
continuity equation, where the vector potential ${\bf A}$
must be included to obtain a gauge invariant expression. 
In a system with a spin-rotational Hamiltonian charge spin components
along arbitrary quantization axes are conserved separately and we can 
introduce a spin-dependent vector potential, ${\bf A}=(\pm e){\bf
A}_c+{\bf s}_{\alpha}\cdot{\bf A}_{\alpha}$, the spin current is
given by $\vj^{\alpha}=-\partial H/\partial {\bf A}_{\alpha}$.  
With this definition a continuity equation expresses local 
conservation of each Cartesian component of spin.  We retain
the same definition when spin-orbit interactions are included, 
although the continuity equation is now violated because the Hamiltonian 
is spin-dependent.  

\noindent 
\section{ Finite size Kubo formula for Spin-Hall conductivity}

We start from linear response theory, Eq.(1) and Eq.(2).
An elementary calculation leads to the following 
formally exact expression for the static z-spin component spin Hall conductivity,
\bea
 \sigma^z_{xy}=-\frac{i\hbar}{L^2}\sum_{\alpha,\alpha'}\frac{f(E_{\alpha})-f(E_{\alpha'})}{E_{\alpha}-E_{\alpha'}}
\frac{\langle\alpha|j_x^z|\alpha'\rangle\langle\alpha'|j_y|\alpha\rangle}{E_{\alpha}-E_{\alpha'}+ i\eta}.
\label{eq:sigmaSH}
\eea
where $j_i= e\partial H/\partial p_i$, $j^z_y=\{\partial H/\partial p_y, s_z\}/2$, with $s_z$ being
$(\hbar/2)\sigma_z$ for electrons and $(3\hbar/2)\sigma_z$ for holes.
In Eq.~(\ref{eq:sigmaSH}) $i \eta$ can be regarded as a complex frequency continued from the real axis to the 
imaginary axis and can be interpreted as an electric-field turn on time.
In metallic systems, like the ones considered here, $\eta$ must exceed the simulation cell level spacing
$\delta E$ in order to obtain bulk values of the transport coefficients considered.
At the same time, $\eta$ must be smaller than all other intensive energy
scales such as the Fermi energy $E_F$, the spin-orbit coupling splitting $\Delta_{SO}$, and
the disorder broadening $\hbar/\tau$, where $\tau$ is the scattering time. 
The finite value of $\eta$ represents the coupling of a finite subsystem of a 
macroscopic conducting sample to its environment, leading for metallic systems to
the loss of resolution of the discrete individual energy levels of the subsystem.
For a finite system with periodic boundary conditions,
the spin-Hall conductivity is a function of $\delta E/E_F,
\eta/E_F,\Delta_{SO}/E_F$, and $\hbar/\tau E_F$.  The macroscopic 
{\em dc} spin Hall conductivity is obtained by extrapolating finite size results 
first to $\delta E \to 0$ ($L \to \infty$) and then to $\eta \to 0$.

Numerical evaluation of the spin Hall conductivity is complicated by the 
substantial fluctuations in finite size system values when $\eta$ is 
small. Following the seminal arguments of Thouless and Kirkpatrick \cite{thouless},
the physically appropriate value for $\eta$ is 
$\eta \sim g\delta E$ where $g=2 E_F\tau$ is the Thouless dimensionless conductance.
The values of $\tau$ quoted in our results were calculated from the 
golden-rule expression for the transport scattering rate,
\bea
\hbar/\tau=2\pi\sum_{\bf k}|V({\bf k-k'})|^2(1-\hat{\bf k}\cdot\hat{\bf
k}')\delta(E_{{\bf k}'}-E_F),
\eea 
which determines the Drude longitudinal charge conductivity via 
$\sigma_D=n e^2 \tau/m=2E_F\tau(e^2/h)$.
The variance of numerical spin Hall conductivities does appear to 
get smaller with system size, to the extent that this trend can be judged
from our numerical results, but relative fluctuations in magnitude are 
still larger than one at small $\eta$ even for the largest system sizes that 
we are able to study.  
In our calculations, the disorder averaged spin Hall conductivity always has the 
same sign as the intrinsic spin Hall conductivity $\sigma^z_{xy}$, negative
for the k-linear Rashba model, and positive for the k-cubic Rashba model.
In Fig.(1) we have chosen a sign convention in which the intrinsic spin-Hall conductivity
is defined as positive.  
\noindent
\section{Spin Hall conductivity Numerical Results}
\subsection{k$^1$ Rashba model}

\begin{figure}[!t]
\begin{center}
\includegraphics[width=0.45\textwidth]{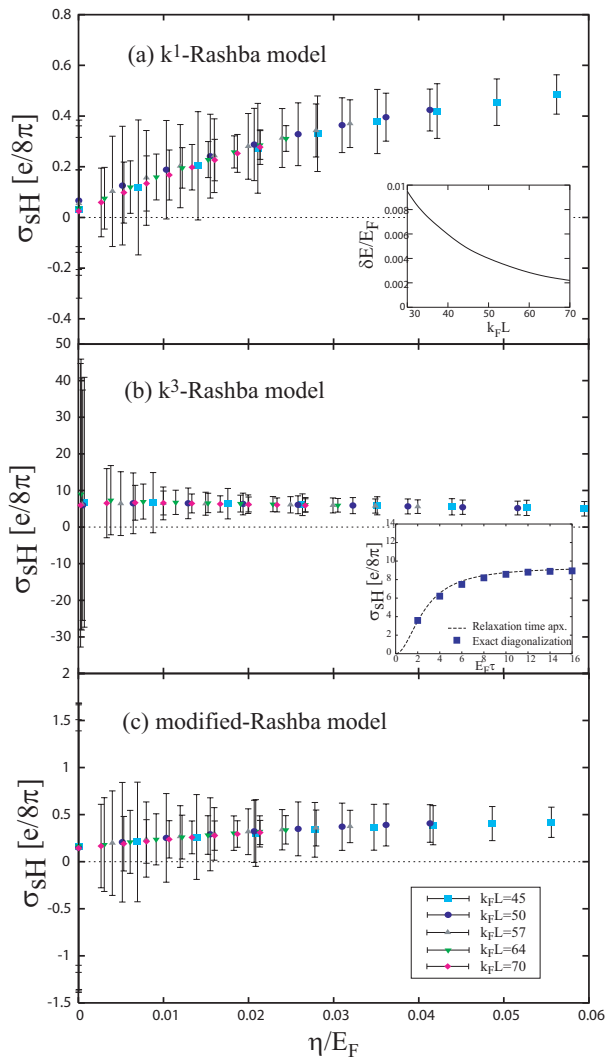}
\caption{
 Spin Hall conductivity for a variety of system sizes as a function of $\eta$ in the k-linear Rashba
 model (top panel), k-cubic Rashba model (middle panel) and modified
 Rashba model (bottom panel).  All of these results are for the case $E_F\tau/\hbar=8$ and 
 $\lambda k_F/E_F=0.2$
 Inset in panel a): Level spacing as a function of system size.
 The results shown in this figure correspond to level spacings varying between $\sim 0.003 E_F$ and 
 $\sim 0.005 E_F$.  Inset in panel b): Disorder strength dependence of the spin Hall conductivity 
 for the $k^3$ Rashba model.  
 The {\em dc} (${\rm Lim}_{\eta \to 0}{\rm Lim}_{\delta E \to 0} \sigma_{SH}(\eta,\delta E)$)
 spin Hall conductivity appears  to vanish for the k-linear Rashba model case only.  
}
\label{fig1}
\end{center}
\end{figure}

In Fig.\ref{fig1} we plot spin-Hall conductivities evaluated for models with 
$E_F\tau/\hbar=8$ with $\lambda k_F/E_F=0.2$ and various 
system sizes as a function of $\eta$ in the k-linear Rashba model (top panel),
the k-cubic Rashba model (middle panel) and the modified Rashba model
(bottom panel). We note that in the regime where $\eta$ is smaller than $E_F,\Delta_{SO}$ and 
$\hbar/\tau$ but larger than
$\delta E$ (plotted as a function of system size as an inset in Fig.\ref{fig1}),
the spin Hall conductivity sometimes changes as a function of $\eta$, strongly so in the  
k-linear Rashba model case. Taking the limit $\eta\rightarrow 0$, extrapolating from the regime where $\eta > \delta E$ 
is satisfied, we find that for the linear Rashba model the spin Hall
conductivity is strongly suppressed in the thermodynamic limit.  Our numerical 
results are consistent with the conclusion from analytic calculations that $\sigma_{sH}$ 
vanishes for this model.
The disorder strength dependence of the spin-Hall conductivity for the k-linear Rashba model is 
shown in Fig.3 where $\sigma_{SH}$ is plotted as a function of $\eta\tau$
and $\eta/E_F$ (inset) fixing $\tau E_F$ at 10, 8, 6, and 4.
These results are consistent with the analytic theory conclusion that 
$\sigma_{SH}$ approaches the intrinsic value for $\omega > \tau^{-1}$,
but that it vanishes for this model for $\omega \to 0$.  In perturbation theory,
the contribution to $\sigma_{SH}$ that varies on the frequency scale $\tau^{-1}$ 
comes from vertex corrections to the intrinsic interband response.
Our previous numerical results which reached an incorrect conclusion on the 
{\em dc} value of $\sigma_{SH}$ for this model, were performed at a 
value of $\eta$ which gives accurate values for the {\em dc} longitudinal 
conductivity but, as we have now learned by extrapolating $\eta \to 0$, not for 
the spin Hall conductivity.

\begin{figure}[!t]
\begin{center}
\includegraphics[width=0.46\textwidth]{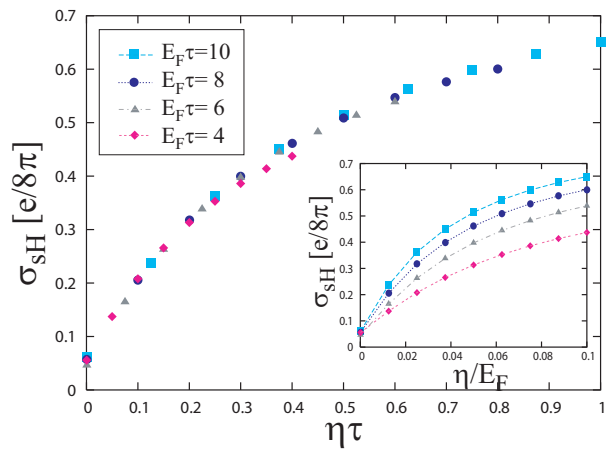}
\caption{
 Spin Hall conductivity as a function of the relaxation time $\tau$ in
 the k-linear Rashba model plotted as a function of $\eta \tau$.
 Inset: Same quantity plotted as function of disorder strength $E_F\tau=10,8,6,4$.
 These results demonstrate that the spin Hall conductivity has a contribution 
 with frequency dependence on the scale $\tau^{-1}$, and are consistent with the 
 conclusion tha the {\em dc} spin Hall conductivity vanishes. 
}
\label{}
\end{center}
\end{figure}

\subsection{k$^3$ Rashba model}

For the Rashba 2DHG the spin Hall conductivity is insensitive to $\eta$; there is 
no evidence of a relatively contribution that has frequency 
dependence on the scale of $\tau^{-1}$.  This finding indicates that 
vertex corrections to the intrinsic spin Hall conductivity, at least for the 
short-range disorder scattering model we have studied, are absent or 
not notably large.  We find similar behavior over a wide parameter range
of $\lambda$ and $\tau$. This finding is consistent with analytical studies of the spin Hall
conductivity by Bernevig and Zhang \cite{bernevig} who find the vertex corrections
vanish in this model. 
Our results should be compared with those from other recent 
numerical studies of mesoscopic
spin transport based on lattice versions
of the Luttinger model for p-doped semiconductors. In both 
quantum well \cite{shehole1} and bulk cases\cite{shehole2}, the intrinsic 
spin Hall conductance is found to be 
robust against disorder in agreement with the present result.
Also, as shown in the inset, the exact-diagonalization numerical results are in very good agreement
with analytic calculations which correct for disorder only by including finite-lifetime 
corrections to the Greens functions that appear in the intrinsic diagram.  (This approximation 
is referred to in the figure as the relaxation time approximation.) 
This implies that the intrinsic effect will be dominant in systems with strong
spin-orbit coupling satisfying $\lambda k_F^3 \gg \hbar \tau^{-1}$.

\subsection{Modified Rashba model}

The modified Rashba model has properties 
intermediate between those of the k$^1$ Rashba and k$^3$ Rashba models since 
the energy spectrum is identical to that of the k$^3$ model, but the
eigenstates are the same as those of the k$^1$ model. 
In this case the $\eta$ dependence of the spin Hall conductivity is shown in Fig. \ref{fig1} (c).
We find a smooth deviation of $\sigma_{sH}$ at small $\eta$ region.
The numerical calculations do find indications of dependence on 
$\eta \tau$, implying that vertex corrections to the intrinsic spin Hall
conductivity do not vanish.  However, the overall spin Hall
conductivity remains finite in the limit $\eta \to 0$. 

No analytic studies of vertex corrections in this model have 
been reported to date, although related models\cite{murakami2} have been
studied.  Using a formalism similar to that developed in
Ref. [\onlinecite{bernevig}], we evaluate the vertex correction to the
renormalized current.
In the limit of $\Delta_{so}\tau\gg 1$ the vertex correction can be
expressed in terms of $\delta
j_{\nu} \equiv J_{\nu}-j_{\nu}\equiv \delta j_{\nu}^i\sigma_i$.  For short
range scatterers,
\bea
 \delta j_{\nu}^i(z,z')=\frac{1}{2\pi\tau\nu_FL^2}\sum_{\vk}{\rm tr}\left[\sigma_iG(\vk,z)j_{\nu}G(\vk,z')\right],
\eea
where $\nu_F$ is the density of states in the absence of spin-orbit
coupling.  After a straitforward calculation, we obtain the following expression
$
 \delta j_{\nu}^i=-\epsilon_{\nu
 i}\frac{1}{2m\nu_F}[\nu_{F+}k_{F+}-\nu_{F-}k_{F-}]-(3/2\nu_F\lambda)\epsilon_{\nu
 i}[\nu_{F+}k_{F+}^2+\nu_{F-}k_{F-}^2].
$
where $k_{F\pm}$ and $\nu_{F\pm}$ are the Fermi wave length and the
density of state in the $\pm$ band.
The spin-dependent Fermi wavelenghts and densities-of-states that appear above 
are given by 
\be
\nu_{F\pm}=\nu_F(1\pm 3m\lambda k_F)^{-1}
\ee
and 
\be
k_{F+}-k_{F-}=-\frac{1}{2m\lambda}(1-\sqrt{1-8(m\lambda k_F)^2}).
\ee
In the weak spin-orbit interaction limit, $\Delta_{so}/E_F<<1$, these may
be approximated by $\nu_{F\pm}\rightarrow \nu_F(1\mp 3m\lambda k_F)$ and
$k_{F+}-k_{F-}\rightarrow -2m\lambda k_F^2$, and then vertex correction
ends up being $\delta \vj=-\lambda k_F^2[\vz\times\vs]$ which ends up canceling
the intrinsic contribution to the spin-Hall conductivity. 
On the other hand, beyond the small spin-orbit coupling limit
the exact cancellation between intrinsic and vertex contributions
does not take place,
consistent with the above numerical result.
This is in sharp contrast with the k-linear Rashba model where the cancellation appears to 
hold for arbitrarily strong spin-orbit coupling.  We comment further on this special 
property of the k-linear Rashba model later in the paper.  

\noindent 
\section{Spectral representation of the SHE and sum rules}
The finite-size Kubo formula 
for the spin-Hall conductivity, Eq.(\ref{eq:sigmaSH}) may be expressed in the form
\bea
 \sigma_{sH}=\frac{\hbar}{L^2}\sum_{\alpha,\alpha'}[f(E_{\alpha})-f(E_{\alpha'})]
\frac{{\rm Im}[\langle\alpha|j_y^z|\alpha'\rangle\langle\alpha'|j_x|\alpha\rangle]}{(E_{\alpha}-E_{\alpha'})^2 + \eta^2}.
\label{eq:sigmaSHR}
\eea
This form is based partly on our finding that the dissipative contribution to the spin Hall conductivity,
which is not included in the above expression is vanishing,  
in analogous with the case of a charge Hall conductivity, the dissipative term strictly vanishes 
when spatial invariance is recovered by averaging over disorder realizations.

It is instructive to consider the following spectral decomposition of the spin-Hall conductivity,
\begin{equation}
 \sigma_{sH} =\int_{0}^{\infty} dE\frac{N(E)}{E^2+\eta^2} 
\end{equation}
where
\bea
N(E) &=&
\frac{2}{L^2}\sum_{\alpha,\alpha'}[f(E_{\alpha})-f(E_{\alpha'})]
{\rm
Im}[\langle\alpha|j_y^z|\alpha'\rangle\langle\alpha'|j_x|\alpha\rangle] \nonumber \\ 
&& 
\delta(E-E_{\alpha}+E_{\alpha'}).
\label{eq:nomura}
\eea
In the following we first focus on the ordinary k-linear Rashba model.
$\sigma_{sH}$ depends on both the phase and the magnitude of the 
matrix elements in Eq.(\ref{eq:sigmaSH}) and on the energy differences of the levels involved.  The size of the 
matrix elements is characterized by the integral of $N(E)$ over all energies which satisfies the following sum rule: 
\bea
\int_0^{\infty}dE N(E) &=& \frac{\hbar}{L^2} \sum_{\alpha} f(E_{\alpha})  
{\rm Im}[\langle\alpha|[j_y^z,j_x]|\alpha\rangle] \nonumber \\
&=& \frac{-e\hbar\lambda}{m L^2} \sum_{\alpha} f(E_{\alpha})  
\langle\alpha| \pi_y \sigma^{x} |\alpha\rangle \nonumber \\ 
&=& \frac{-e \hbar^2 \langle H_{so} \rangle}{2 m L^2} \equiv M^{0}.
\label{sumrule} 
\eea
The final form for the zeroth moment of the Hall spectral function ($M^{0}$) 
in Eq.(\ref{sumrule}) follows from the observation
that the two terms in the Rashba spin-orbit interactions must have identical expectation values 
if isotropy is recovered in the thermodynamic limit.
For $\hbar/\tau \ll \Delta_{so}$, $\langle H_{so} \rangle$ is close to its 
value in the perfect crystal state.  We note that 
this expression is valid both in the presence (see below) and absence of an external 
magnetic field and that $\pi_y$ is the kinetic momentum in the $\hat y$ direction.

The left panel in Fig. 4 shows the spectral function $N(E)$ as a function
of $E$ at $\lambda k_F/E_F=0.2$ and at two disorder strength
$E_F\tau=10$ and $E_F\tau = 4$. N(E) has a positive peak at $E$ corresponding with the
spin-orbit splitting $2\lambda k_F$ ($k_{F+}$ and $k_{F-}$ are approximately equal)
and becomes negative with very small magnitude at small $E$.  The negative 
contribution at small $E$ corresponds to the vertex correction contribution to the 
spin Hall conductivity. The large peak near the spin orbit splitting energy corresponds 
to the intrinsic contribution to the spin Hall 
effect. In the limit of small spin-orbit coupling, the energetic width of the 
interband peak vanishes and the intrinsic spin-Hall conductivity is proportional to the ratio
of the sum rule and the square of the spin-orbit splitting. 
The vertex correction contribution is enhanced in $\sigma_{SH}$ by small
energy denominators. 

As first emphasized by Rashba\cite{rashba2}, insight into the 
spin Hall conductivity of the linear Rashba model can be achieved by 
introducing an external magnetic field.  Letting
\bea
\hbar\vk\rightarrow -i\hbar\nab+e{\bf A}(\rr),
\eea
we introduce a magnetic field $B$ perpendicular to the plane.
(${\bf A}(\rr)=By\hat{\bf x}$.)
The charge current and the spin current in this case are given by
$\vj=-e(i/\hbar)[H,\rr]=-e(\vpi/m-\lambda\vz\times\vs)$ and $\vj^z=\{\vpi/m,\sigma_z/2\}/2$, where
$\vpi=-i\hbar\nab+e{\bf A}(\bf r)$ is the kinetic momentum.
We consider the case of Landau level filling factor $\nu\simeq 7$ and $\lambda/l_B\bar\omega_c=2$ as an example in the
following. The energy spectra are linear in the Landau level index in
both bands and the spin-orbit splitting near the Fermi level is approximately 
$5\hbar\omega_c$ as shown in right bottom panel in Fig. 4.
We find that the spectral function $N(E)$ has several peaks. The two
left most peaks can be identified as the intra-band contributions
that evolve into the vertex correction at zero field while  
the rest correspond to the inter-band contributions that give the intrinsic 
spin Hall effect.  Interestingly the intraband contribution has both positive 
and negative peaks, with the negative peak appearing at lower energy and 
therefore having a larger contribution to the spin-Hall conductivity.
As pointed out by Rashba\cite{rashba2} the intraband and interaband contributions to
$\sigma_{SH}$ cancel, as in the zero field case with disorder discussed above.

\begin{figure}[!t]
\begin{center}
\includegraphics[width=0.46\textwidth]{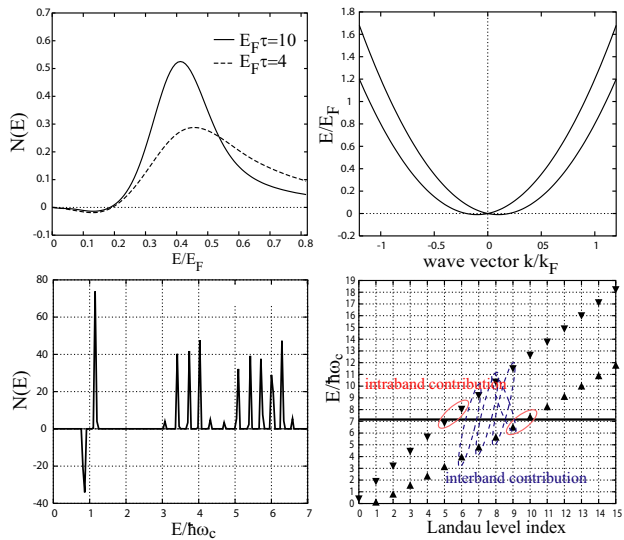}
\caption{
The spin Hall spectral function $N(E)$ in the absence (top left panel) and presence
(bottom left panel) of a magnetic field.
The right panels show the corresponding single-electron energy spectra
in the absence of disorder.}
\label{}
\end{center}
\end{figure}

\noindent
\section{What is special about the k-linear Rashba model?}

Perturbation theory calculations and these numerical calculations consistently 
indicate that the k-linear Rashba model has zero spin Hall conductivity for any
scalar random potential.  This property is special to the k-linear model and 
is not generic, as indicated for example by our numerical results for
the other models considered in this paper.  How should we understand this 
exceptional behavior?  One possibility is to explain it in terms of vertex 
corrections to the model's current operator, which cancel spin-dependent
contributions on the Fermi surface.   This explanation is not fully satisfactory,
however, since the spin-Hall conductivity depends partly on contributions 
away from the Fermi surface, and the same cancellation does not cause the 
charge Hall conductivity of corresponding ferromagnet models to vanish.  
Instead, as discussed previously elsewhere in the literature\cite{dimitrova,loss},
the source of the special behavior is almost certainly the following 
exact relationship between the time dependence of the total spin operator and 
the spin-current operator:
\begin{equation}
\dot{s}_{\,x}=i[H_R,s_{x}]/\hbar = \frac{2m \lambda}{\hbar^2}  j_x^{z}.
\label{eq:sdot} 
\end{equation} 
(This type of relationship between in-plane spin equation of motion 
and in-plane spin-current, can be generalized to any model with spin-orbit
coupling that is linear in momentum, so that most of the conclusions reached below
to the 2D Dresselhaus model and even to models that include 
both Dresselhaus and Rashba interactions and do not have circular Fermi lines.) 
Since the spin-density must approach a constant in the steady state, it follows that 
the non-equilibrium steady state expectation value of the left-hand-side of this equation,
and hence also of the right hand side, must vanish.  We believe that this argument is 
essentially valid, although there is subtlety in its application because the 
conductivity is defined by taking the thermodynamic limit and then the {\em dc} limit.
(The same argument, naively applied, could be used to prove that the 
drift velocity vanishes in the steady-state induced by an external electric field.) 
Below we discuss the implications of this identity for the finite-size 
spin Hall conductivity calculations, and then discuss some of its implications 
for semiclassical descriptions of spin transport in the Rashba
model and other systems. 

\subsection{Finite-size Kubo formula}

We start by considering the linear response of an in-plane spin component 
to a constant change in the vector potential in a finite-size system.  The 
vector potential change gives rise to a perturbation $H'=-\vj\cdot\delta {\bf A}$: 
\bea
\frac{\delta S_{\mu}}{\delta A_{\nu}}(\omega)&=&-\frac{i}{\hbar}\int_0^{\infty}dt\langle[s_{\mu}(t),j_{\nu}]\rangle
e^{i(\omega+i\eta)t}/L^2  \nonumber \\
&=&-\frac{1}{ L^2}\sum_{n,n'}\frac{f(E_n)-f(E_n')}{E_n-E_n'+\omega+i\eta}\langle
n|s_{\mu}|n'\rangle\langle n'|j_{\nu}|n\rangle.  \nonumber \\ 
\label{dsda}
\eea
For a square finite size system with side $L$,
a change in vector potential by $\delta A_{\nu}$ 
corresponds to a change in the boundary condition phase 
by $ 2 \pi L \delta A_{\nu}/ \Phi_0$ where $\Phi_0$ is the electron flux quantum.
Using the relation between the time dependence of the spin operator and 
the spin-current operator, it follows that
\begin{equation} 
(i/\hbar)(E_n-E_{n'})\langle
 n|s_{\mu}|n'\rangle=-2m\lambda\langle n|j^z_{\mu}|n'\rangle/\hbar^2.
\end{equation} 
Comparing with Eq.~(\ref{eq:sigmaSH})
 we find that 
\bea
 \sigma_{\mu\nu}^z(z)=\frac{\hbar^2}{2m\lambda}\frac{\delta S_{\mu}}{\delta A_{\nu}}(z).
\label{eq:sigmaR2}
\eea
where $z = \omega + i \eta$ is a complex frequency.  This is an exact
 expression for the k-linear Rashba model with arbitrary scalar impurities.
For a disordered R2DES $s_{\,x}$ in general has a boundary condition and disorder potential
dependent expectation value $\propto L$,
corresponding therefore to a spin-density per unit volume that vanishes
in the thermodynamic limit.
Because the $z \to 0$ value of the spin density response to vector potential 
is proportional to the derivative of the ground state spin density with respect 
to boundary condition phase angle in the direction of the vector potential, 
\begin{equation}
\lim_{\eta \to 0} \sigma_{sH} = \frac{ \hbar^2}{2m \lambda L} \frac{\partial S_{\,x}}{\partial \phi_y},
\label{eq:relationtosx}
\end{equation} 
it is evident that the average of the $z \to 0$ value of this response function over boundary 
conditions is zero.  (Since the  spin-density must be a periodic function of 
$\phi_y$ with period $2 \pi$, the integral of its derivative over any period must 
vanish.  In Eq.(\ref{eq:relationtosx}) $\phi_y$ is the boundary condition phase angle in the 
$y$-direction and $S_x={\rm Tr}[\rho \frac{\sigma_x}{2}]$.) This appears to be the conclusion that can most confidently be
drawn about Kubo formula properties from Eq.~(\ref{eq:sdot}). 
For $\eta =0$, the typical value of the spin Hall conductivity at a particular 
boundary condition is large in magnitude, indicative of large persistent spin currents 
in finite size systems.  Our numerical results for the spin-Hall
conductivity appear to be consistent with the natural {\em ansatz} that averaging
over boundary conditions is equivalent to averaging over disorder realizations in
finite-system calculations of the spin-Hall conductivity.  The fact that these 
averages at $\eta =0$ appear to yield the same values for the spin-Hall conductivity 
as extrapolations from $\eta > \delta E$, guarantees that the equation of 
motion argument for vanishing spin Hall conductivity in the linear Rashba model is valid.  

We note that it is possible to establish that the spin-Hall conductivity vanishes at integer
Landau level filling factors in the absence of disorder
without appealing to a concrete calculation by using total spin equation of motion identities.  
In this case, the boundary condition phase angle $\phi_y$ just corresponds to an $x$-direction
guiding center shift in a translationally invariant system.  In sharp contrast to the 
zero field case, there is no $\phi_y$ dependence in the absence of disorder; the result for 
any boundary condition equals the zero result obtained by averaging over boundary conditions.
Because of the gap between Landau levels, no subtleties arise in taking the thermodynamic limit.
The spin Hall conductivity clearly vanishes at any integer filling factor.

There literature contains some arguments that the spin Hall conductivity 
vanishes for any model.  For example, Sheng et al.\cite{sheng2} have
performed a numerical simulation similar with present work, and
have concluded that the spin Hall effect of source of the 
zero spin Hall conductivity of the linear Rashba model is more 
general.  In particular they argue that because all energy eigenvalues have anti-crossing
behavior as a function of boundary conditions (or equivalently flux
$\Phi$ through a cylindrically shaped sample) spin transports cannot arise.
We note that the linear response regime attains with external
field $eEL=-e\ d\Phi/dt$ small compared with all relevant energy scales but larger
than the level spacing $\delta E$ to generate Landau-Zener tunneling
through anti-crossing gaps. Consequently the adiabatic argument of
Ref.[\onlinecite{sheng2}] cannot capture the linear response of the spin transport.
We rather conclude that strong suppression of the spin Hall conductivity is 
an accidental property of particular models, not a generic effect. As we have seen in this article
the k-cubic model gives a good example of a model for which the 
intrinsic spin Hall effect is dominant.  The spin Hall effect observed in this 
model can be understood as an intrinsic effect\cite{wunderlich,bernevig}.
We note that the spin Hall effect is observed as spin accumulation.  Although
there is no analytic theory of spin accumulation due to an intrinsic spin Hall
effect, we note that numerical studies of spin accumulation\cite{nikolic2,nomura2}
show little accumulation in the k-linear Rashba model and robust 
accumulation in the k-cubic model. These results are consistent with
the present spin transport study and a naive theory of spin accumulation.

\subsection{Implication for Semiclassical Theory of Coupled Charge and Spin Transport}

From the equation of motion for the density matrix it follows quite generally 
that the time derivative of the spin-density can be related to the equation of 
motion for the averaged spin-density:
\begin{equation}
\frac{d S_{\mu}}{d t} = \frac{d}{d t}{\rm Tr}\left[\rho(t) \frac{\sigma_{\mu}}{2}\right]. 
\end{equation}
In the steady state which balances acceleration by an external electric field
with disorder scattering, the density matrix $\rho$ is constant and the 
spin-density is expected to saturate at a finite value.  As discussed above,
it follows from this argument that $J^z_{\mu} 
\propto (d/dt){\rm Tr}[\rho \sigma_{\mu}/2 ]$ vanishes.  It is interesting to 
compare the single-band semiclassical theory of spin
  transport\cite{culcer} with this result.
  In a homogeneous system this theory describes the spin-density
dynamics by the following equation:
\bea
\frac{d S_{\mu}}{d t}
&=& \frac{1}{L^2}\sum_{\vk}{\rm tr}\left[
f{\dot{s}_{\mu}}+\left(\frac{d f}{d t}\right)s_{\mu} \right]\nonumber \\
&=&
- \frac{2m\lambda}{\hbar^2}J^{z,int}_{\mu} -\frac{S_{\mu}}{\tau}.
\label{dsdt}
\eea
The first term in the first form for the right hand side corresponds to the spin torque 
term which describes spin-density dynamics in the absence of collisions.  The effect 
of collisions which scatter electrons between Bloch states is
accounted for by the second term.\cite{culcer}  In the final form for the right 
hand side we have introduced the relaxation time approximation for the scattering term
and recognized that the collision free expression for the spin-density evolution 
is proportional to the Hall spin-current in the absence of collisions and hence to
the intrinsic spin Hall conductivity.  
In the strong spin-orbit scattering limit on which we focus, the 
spin relaxation time\cite{dyakonov,halperin} that appears 
in this equation may be approximately identified with the momentum relaxation 
time.  The steady state in-plane spin-density induced by an electric field
is proportional to the intrinsic spin Hall conductivity and to the momentum
relation time $\tau$.  We know that above semiclassical argument of the spin-Hall 
conductivity fails to capture vertex corrections, because it does not 
properly account for the influence of disorder on the interband 
components of the density matrix response.\cite{sinitsyn}  Since
vertex corrections also change the value of the in-plane spin induced by 
an electric field,\cite{inoue} this theory may also fail to account 
quantitatively for the value of in-plane spin-density induced by an external 
electric field.  
 
\section{Summary}

In this article, we have studied spin transport driven by an
electrical potential bias in two-dimensional electron and hole systems with spin-orbit 
coupling due to structural inversion asymmetry, primarily using 
finite-size exact diagonalization as a tool.  We have studied three different models of spin orbit coupling, 
the standard k-linear Rashba model, a k-cubic Rashba model appropriate for two-dimensional
hole systems in narrow quantum wells, and a modified k-linear Rashba model with 
eigenspinors like that of the standard Rashba model and eigenvalues like that of 
the k-cubic Rashba model.  In these systems a current of spins oriented perpendicular to the 
two-dimensional layer flows perpendicular to the direction 
of the electric field, an effect known as the spin Hall effect.
For the models we have studied the expectation value of the spin current in
the perfect crystal Bloch eigenstates is zero, implying that there is no
skew-scattering-induced Bloch state occupation number change contribution
to the spin Hall effect.  The spin Hall effect is entirely due to interband coherence
induced in the system by the electric field.  When disorder is treated perturbatively 
the spin Hall effect can be separated into an intrinsic contribution that is a property
of the perfect crystal electronic structure along, and a disorder-related vertex correction
contribution that remains finite even when the scattering rate vanishes. 
This vertex-correction is partially analogous to the scattering angle weighting 
correction that vertex corrections introduce into the theory of the longitudinal
conductivity.  

The three models we study differ qualitatively on how vertex corrections 
alter the intrinsic spin Hall effect.  We evaulate the spin Hall conductivity 
numerically for a finite system and at a finite frequency $i \eta$ continued
to the imaginary axis. The frequency $\eta$ can be thought of as a turn 
on rate for an electric field, or as energy level broadening due to the 
coupling of the small system being studied numerically to the rest of a 
macroscopic sample.  The thermodynamic limit {\em dc} spin Hall conductivity 
must be calculated by first letting the system size become large and 
then letting $\eta \to 0$; the vertex correction appears as a dependence
of the spin Hall conductivity on $\eta \tau$.  
In the k-linear Rashba model we find that the spin Hall 
conductivity depends strongly on $\eta \tau$, vanishing for 
$\eta \to 0$.  This finding is consistent with analytical calculations have shown that
the vertex correction strongly suppresses the intrinsic contribution to
the spin Hall conductivity for this model.  For the k-cubic Rashba model we 
find that vertex corrections vanish, a finding that may hold only for the
short-range disorder scattering model we apply.  For the modified k-linear 
Rashba model, the vertex corrections do not vanish and alter the 
intrinsic spin Hall effect by a fraction that decreases with increasing 
spin-orbit coupling strength.  Taken together, these results demonstrate that 
the intrinsic interband spin Hall conductivity can be altered by vertex corrections,
depending on details of the electronic structure and the disorder potential.
The special situation which leads to a vanishing total spin Hall conductivity for the 
k-linear Rashba model is related to the relationship between spin equations-of-motion 
and spin-currents that applies only for systems with spin-orbit coupling that is 
linear in momentum.  
Since the spin Hall conductivity of the k-cubic model is purely interband, and vertex 
corrections are weak for this model, we conclude that the spin-Hall-induced 
spin accumulation observed in a two-dimensional hole gas by Wunderlich {\em et al.}\cite{wunderlich} 
(for which the k-cubic model is applicable), must be due primarily to the 
intrinsic spin Hall effect.   

\section*{Acknowledgement}

The authors thank D. Culcer, T. Jungwirth, M. Koshino, S. Murakami, 
N. Nagaosa, Q. Niu, S. Onoda, E.I. Rashba, D. N. Sheng for useful discussions.  One of the authors K.N. is supported by 
the Japan Society for the Promotion of Science by a  
Research Fellowship for Young Scientists.  This work has been supported by 
the Welch Foundation and by the Department of Energy under grant DE-FG03-02ER45958.

\end{document}